\documentclass[11pt]{article}
\usepackage{amsmath,amssymb}
\usepackage[left=2cm,right=2cm,top=2.5cm,bottom=2.5cm]{geometry}
\usepackage{float}
\usepackage{graphicx,epstopdf}
\graphicspath{{figures/}}
\usepackage{caption,subcaption}
\usepackage{listings}
\usepackage{url}
\usepackage{soul, color}
\usepackage{xcolor}
\usepackage{bm}
\usepackage[colorlinks=true,linkcolor=blue]{hyperref}
\usepackage[export]{adjustbox}
\usepackage{verbatim}
\usepackage{siunitx}
\usepackage{stackengine}
\usepackage{comment}

\usepackage{parskip}
\usepackage{textcomp}
\usepackage{mathtools}
\usepackage[normalem]{ulem}
\usepackage{wrapfig} 
\usepackage{setspace} 
\title{Mathematical Modeling Insights into Improving CAR T cell Therapy for Solid Tumors: Antigen Heterogeneity and Bystander Effects }

\author{Erdi Kara\thanks{Corresponding Author, \url{erdikara@spelman.edu }},\thanks{Department of Mathematics, Spelman College}, 
T. L. Jackson, \thanks{Department of Mathematics, University of Michigan}, 
Chartese Jones, \thanks{Department of Mathematics, University of Missouri},
Reginald L. McGee II, \thanks{Department of Mathematics, College of the Holy Cross}
Rockford Sison, \thanks{Department of Mathematics, Spelman College}
}

\date{}

\begin{document}
\maketitle

\begin{abstract}


As an adoptive cellular therapy, Chimeric Antigen Receptor T-cell (CAR T-cell) therapy has shown remarkable success in hematological malignancies, but only limited efficacy against solid tumors. Compared with blood cancers, solid tumors present a unique set of challenges that ultimately neutralize the function of CAR T-cells. One such barrier is antigen heterogeneity - variability in the expression of the antigen on tumor cells. Success of CAR T-cell therapy in solid tumors is unlikely unless almost all the tumor cells express the specific antigen that CAR T-cells target. 
A critical question for solving the heterogeneity problem is whether CAR T therapy induces bystander effects, such as antigen spreading. Antigen spreading occurs when CAR T-cells activate other endogenous antitumor CD8 T cells against antigens that were not originally targeted. In this work, we develop a mathematical model of CAR T-cell therapy for solid tumors that takes into consideration both antigen heterogeneity and bystander effects. Our model is based on in vivo treatment data that includes a mixture of target antigen-positive and target antigen-negative tumor cells. We use our model to simulate large cohorts of virtual patients to gain a better understanding of the relationship between bystander killing. We also investigate several strategies for enhancing the bystander effect and thus increasing the overall efficacy of CAR T-cell therapy for solid tumor.
\end{abstract}

\newpage


\section{Introduction}
\label{sec:intro}

Chimeric antigen receptor T‐cell (CAR‐T) therapies are one of the most innovative cancer treatment modalities available today and have radically altered the treatment landscape for multiple malignancies \cite{sterner2021car}.  Chimeric antigen receptors (CARs) are engineered fusion proteins that allow  T cells to recognize specific antigens expressed on cancer cells  \cite{cappell2023long}. By removing immune cells, manipulating their genes, and implanting them back into the body, scientists have created a way to ``teach" our immune system to identify and destroy cancer cells. This procedure got its first two FDA approvals in 2017 \cite{BRAENDSTRUP202057}. Since then, four more therapies have been approved\cite{ApprovedTherapies,cappell2023long}. These therapies have demonstrated repeated success in handling liquid tumors including certain leukemias, lymphomas, and most recently, myeloma \cite{safarzadeh2021car}. Given the promising clinical responses for several hematological malignancies, there is a vested interest in examining the applicability to solid tumors.  Unfortunately, CAR T-cell therapy has not generated even close to the same level of success in the fight against solid tumors due to several challenges that are often interconnected \cite{safarzadeh2022recent}. Some difficulties in applying CAR T-cell therapy to solid tumors are tumor antigen heterogeneity, trafficking and infiltration into tumor tissue, and immunosuppressive tumor micro-environment\cite{Marofi_2021}. Of these, antigen heterogeneity - variability in the expression of the antigen on tumor cells is particularly challenging  \cite{jia2022heterogeneity}.  The ideal tumor associated antigen (TAA) would be expressed on all tumor cells because CARs can only attack cells having the targeted antigen.  Therefore, antigen heterogeneity impairs the detection of cancer cells by T cells and therapuetic success is unlikely unless almost all the tumor cells express the TAA \cite{marofi2021car}.


A critical question for solving the heterogeneity problem is can CAR T-cell therapy induce "bystander effects" \cite{sanchez2017antigen,upadhyay2021critical}. The term bystander effect refers to the phenomenon where CAR T-cells not only target cancer cells expressing a specific antigen but also stimulate cytotoxic activity against nearby cancer cells that lack the targeted antigen expression. In the same context, bystander cells refer to immune cells triggered by CAR T-cells playing a role in the killing of tumor cells. These bystander cells can include CD8 T cells, innate immune cells (such as macrophages, neutrophils, or natural killer (NK) cells), or even antigen-presenting cells (APCs) involved in antigen cross-presentation. The bystander effect allows CAR T-cells to exert their therapeutic impact beyond the targeted antigen-expressing cells, contributing to the eradication of heterogeneous cancer cell populations commonly found in solid tumors.  Two types of bystander effects are indirect tumor cell killing and  antigen spreading.  Indirect killing could result from activation macrophages and/or natural killer cells by cytokines released after CAR engagement.  Antigen spreading occurs when CAR T-cells induce the activation of other endogenous antitumor CD8 T cells against other antigens that were not originally targeted \cite{upadhyay2021critical}. It usually occurs following initial therapy-mediated tumor destruction, which leads to the release of secondary tumor antigens.  The extent to which this occurs will be critical in setting the thresholds of antigen expression that will be used to define eligible candidates.

Experimentally evaluating the bystander effect in CAR T-cell therapy presents several challenges. Firstly, relevant experimental models are required to accurately assess immunologic bystander effects. While many preclinical studies have utilized immunodeficient mouse models with human tumors, assessing the bystander effect necessitates intact immune systems in mouse models. Intact immune system is essential to evaluate the extent to which CAR T-cells can effectively cure antigen-positive and target antigen-negative tumor types and thus define the presence and impact of bystander effects. Secondly, defining and quantifying bystander effects can be technically challenging. Precise characterization of antigen expression profiles on tumor cells, designing well-defined mixtures of antigen-positive and antigen-negative tumor cell populations, and accurately measuring the impact of CAR T-cells on these populations require sophisticated techniques and methodologies \cite{klampatsa2020analysis}. 

Increasingly, mathematical models have been used to study CAR T-cell therapy \cite{Sahoo2020,Leon2021,Barros2020}. However, to the best of our knowledge, there is no mathematical model investigating bystander effect of CAR T-cells in solid tumors. In this paper, we build a mathematical model based on an in-vivo experiment \cite{klampatsa2020analysis} that includes a mixture of target antigen-positive and target antigen-negative tumor cells. In our model, we investigate the interaction of four distinct cell populations; antigen-positive tumor cells, antigen-negative tumor cells, CAR T-cells and bystander cells. Our results indicate that augmenting the number of CAR T-cells reaching the tumor site does not enhance the therapeutic outcome, even in cases of high antigen expressivity, suggesting the need for bystander effect . However, our virtual patient analysis reveals that enhancing the cytotoxic capability of bystander cells can generate significantly improved outcomes.

The plan for the paper is as follows. In Sec-\ref{sec:modelling} we describe the mathematical model designed to investigate antigen heterogeneity and bystander effects and explain the key steps of parameter estimation using in-vivo data provided in \cite{klampatsa2020analysis}. 
In section Sec-\ref{sec:results},  we develop a heterogeneous virtual patient population that reflects the variability in the experimental data to simulate different treatment scenarios to gain a better understanding about bystander effect on therapeutic outcome.
Finally, in Sec-\ref{sec:discuss}, we explore the potential extension of our model to address various scenarios regarding the bystander phenomenon.

\section{Model Formation and Calibration}
\label{sec:modelling}

\subsection{Model Equations}
\label{sec:odes}

Our mathematical model investigates the evolution of four distinct cell populations over the course of treatment. Specifically, we consider the total tumor population as consisting of two sub-populations denoted by $T_{s}$ and $T_{r}$, which respectively represent the quantities of antigen-positive and antigen-negative tumor cells, measured in units of $\unit{mm^3}$. Additionally, we denote by $C$ and $B$ the quantities of CAR T-cells and bystander cells present within the tumor microenvironment , also measured in units of $\unit{mm^3}$.

\begin{align}
&\dfrac{dT_s}{dt} = r_1T_s\Big(1-\dfrac{T_s+T_r}{K_1}\Big) - D_{B}T_s - D_{C}T_s \label{eq:Ts} \\  
&\dfrac{dT_r}{dt} = r_2T_r\Big(1-\dfrac{T_s+T_r}{K_1}\Big) - D_{R}T_r \label{eq:Tr}\\ 
&\dfrac{dC}{dt} = v_{C}(t) - \gamma_{C}C - \mu_C\log\Big(\dfrac{B+C}{K_2}\Big)\dfrac{D_C^2}{k+D_C^2}C - \omega_{C}CT_s  \label{eq:C} \\ 
&\dfrac{dB}{dt} = b - \gamma_{B}B - \mu_B\log\Big(\dfrac{B+C}{K_2}\Big)\dfrac{D_B^2}{k+D_B^2}B 
    - \omega_{B}B(T_s + T_r) \label{eq:B}  
\end{align}

\noindent
where 
\begin{subequations}
\label{eq:fracterms}
\begin{align}
    D_B &= d_B\dfrac{(B/T_s)^l}{s+(B/T_s)^l}  \label{subeq:DB}\\
    D_C &= d_C\dfrac{(C/T_s)^l}{s+(C/T_s)^l}  \label{subeq:DC}\\
    D_R &= d_B\dfrac{(B/T_r)^l}{s+(B/T_r)^l} \label{subeq:DR}
\end{align}
\end{subequations}

In the absence of CAR T-cells and bystander cells, our model accounts for tumor growth through competition-based dynamics. The first two terms in equations \eqref{eq:Ts} and \eqref{eq:Tr} respectively represent the logistic growth of the antigen-positive and antigen-negative tumor sub-populations, each towards a carrying capacity $K_1$ with its own growth rate $r_1$ and $r_2$.

In Eq.\eqref{eq:Ts}, the killing of the antigen-positive tumor population is modeled through two ratio-dependent cell-kill terms denoted by $D_B$ and $D_C$. Those terms were originally proposed by de Pillis et al. in \cite{de2005validated} to capture the interaction between tumor cells and immunological components, including CD8 positive T cells and natural killer cells. Therein, they discovered that  tumor cell lysis by 
tumor-specific CD8-T cells is better modelled by using a term that is a function of the ratio of T cells to tumor cells rather than a usual linear cell-kill term. We also tested linear cell-kill and confirmed that it  does not capture the dynamics of  our data and thus we adopt the ratio-dependent functional form for cell kill for our model.

Ratio dependent killing terms are  given in Eq.\eqref{eq:fracterms}. The parameter $l$ describes the degree of dependence of the killing rate on the ratio of CAR T-cells or bystander cells, while the parameter $s$ characterizes the steepness of killing of the corresponding terms. 
The parameters $d_B$ and $d_C$ represent the maximum killing rate that can be achieved when the concentrations of bystander and CAR T-cells are at their respective saturation levels. In Eq.\eqref{eq:Ts}, CAR T-cells selectively recognize and target the antigen-positive tumor cell population, leading to their death. Additionally, the same antigen-positive cell population can be killed by bystander cells in the presence of a bystander effect. In Eq.\eqref{eq:Tr}, the antigen-negative tumor cell population is only inhibited by bystander cells, since these cells are not recognized by CAR T-cells due to the absence of the target antigen.

In the context of the equations \eqref{eq:C} and \eqref{eq:B}, the parameters $\gamma_C$ and $\gamma_B$ represent the natural death rates of CAR T-cells and bystander cells, respectively. It has been observed that CAR-T cells, like T cells responding to chronic viral infections or certain cancers, can experience a loss of proliferative capacity due to prolonged exposure to antigen stimulation. In this sense, the parameter $w_C$ represents the rate of exhaustion of CAR T-cells resulting from exposure to the antigen-positive tumor population, while $w_B$ represents the rate of exhaustion of bystander cells due to their interaction with both tumor cell populations. We assume that the recruitment of CAR T-cells and bystander cells follows the Michaelis-Menten dynamics, as outlined in \cite{kuznetsov1994nonlinear}, which are represented by the terms $\dfrac{D_C^2}{k+D_C^2}$ and $\dfrac{D_B^2}{k+D_B^2}$, respectively. Here, the maximum recruitment of CAR T-cells and bystander cells are denoted by $\mu_C$ and $\mu_B$, respectively. Note that both terms are scaled by the immune cell competition term $\log{\dfrac{B+C}{K_2}}$, which limits the proliferation of immune cell components beyond the carrying capacity $K_2$ \cite{kimmel2019response}. 

CAR T-cell injection $v_C(t)$ is modelled as a one-time increase in the amount of CAR T-cells on the days of injection.

\begin{equation*}
v_{C}(t)=
\begin{cases}
    I_{C}, & \text{if }  t = \text{injection day} \\
    0, & \text{otherwise}
\end{cases}
\end{equation*}

where $I_{C}$ represents the amount of CAR T-cell injected at time day of treatment.

\subsection{Parameter Estimation}
\label{sec:paramest}

We provide the parameter values associated with our model in Table-\ref{table:params}. To determine the key parameters in our model, we utilize data generated in Klampatsa et.al \cite{klampatsa2020analysis} where the bystander effect is investigated in-vivo in a  systematic way. 

As defined in the introduction, bystander effects  refer to the phenomenon where CAR T-cells are able to eliminate tumor cells that do not express the target antigen via a variety of mechanisms including Fas-FasL death receptor engagement and endogenous CD8 T activation \cite{upadhyay2021critical}. Klampatsa et.al describes a mixing model developed using CAR T-cells that react against a human mesothelin-expressing solid murine tumor cell line, allowing for direct testing of the bystander hypothesis. Specifically, in their experiment, $2\times10^6$  human mesothelin (AE17om) were seeded into the right flanks of the mice and tumor population reached to approximately 50 $\unit{mm^3}$ by the end of 3rd day. The treatment scheme is implemented via administration of two doses of $10^7$ transduced M11 CAR T-cells on day 7 and day 9. It was reported that while M11 CAR T-cells could cure tumors that were 100\% mesothelin positive, they could not cure tumor that were 90\%, 75\% or 50\% mesothelin positive, indicating the lack of bystander effect. However, they found that administering
low-dose(100 mg/kg) cyclophosphamide (CTX) to the mice prior to treatment resulted in a bystander
effect and the cure of tumor mixtures.  Each of these experiments included control groups, which means that some subjects were not treated and served as a comparison to the treated group. We will use their data to sequentially calibrate some of our model parameters.

In the subsequent analysis, we used free online tool Plot Digitizer to extract the data provided in the aforementioned publication \cite{PlotDigitizer}. In our model, we work with the amount of cells in \unit{mm^3}, which is more convenient for our purposes than the original paper's use of cell number. To convert between the two units, we assume that $10^6$ cells can be approximated by 1\unit{mm^3} of cells. Parameters that are not discussed below are taken from \cite{owens2021modeling}.

\begin{flushleft}
    \begin{table}[ht!]
    \begin{center}
    \begin{tabular}{ c  p{6.0cm}  p{3cm}  p{2cm} p{2cm}}
    \hline
    \textbf{Parameter} &  \textbf{Description}  &  \textbf{Value}  &  \textbf{Units} & \textbf{Source}  \\
    \hline
    $r_1$ & $T_s$ proliferation rate & 0.151 & \unit{day^{-1}} & fitted \\
    $r_2$ & $T_r$ proliferation rate & 0.180 & \unit{day^{-1}} & fitted \\
    $K_1$ & $T_s$ and $T_r$ carrying capacity & $5.9 \times 10^{3}$ & \unit{mm^{3}} & \cite{owens2021modeling} \\
    $l$ & exponent of tumor lysis by either C or B & 1.291 & unit-less & fitted \\
    $\mu_C$ & max recruitment rate of CARTs by antigen-positive tumor lysis & 0.6 & \unit{day^{-1}} & fitted \\
    $d_C$ & maximum killing rate of antigen-positive cells via CAR-T cells & 0.27 & \unit{day^{-1}} & fitted \\
    $s$ & steepness of fractional antigen-negative tumor kill by bystanders & $3.05 \times 10^{-1}$& unit-less & \cite{owens2021modeling} \\
    $\gamma_C$ & CARTs death rate & $2.93 \times 10^{-2}$ & \unit{day^{-1}} & \cite{owens2021modeling} \\
    $k$ & steepness of cart/bystander recruitment & $2.019 \times 10^{-7}$ & \unit{day^{-2}} & \cite{owens2021modeling} \\
    $\omega_C$ & cart exhaustion due to antigen-positive cells & $3 \times 10^{-5}$ & \unit{mm^{-3}day^{-1}} & \cite{owens2021modeling} \\
    $K_2$ & immune cell carrying capacity & $1.65 \times 10^{3}$ & \unit{mm^3} & \cite{owens2021modeling} \\
    $d_B$ & maximum killing rate of antigen-positive/antigen-negative cells via bystanders & 0.27 & \unit{day^{-1}} & fitted \\
    $\mu_B$ & max recruitment rate of bystanders by antigen-positive tumor lysis & 0.82 & \unit{day^{-1}} & fitted \\
    $b$ & base recruitment rate of bystanders & $5 \times 10^{-2}$ & \unit{mm^3day^{-1}} & \cite{owens2021modeling} \\
    $\gamma_B$ & bystander death rate & $2 \times 10^{-2}$ & \unit{day^{-1}} & \cite{owens2021modeling} \\
    $\omega_B$ & bystander exhaustion due to antigen-positive/antigen-negative cells & $3.42 \times 10^{-6}$ & \unit{mm^3day^{-1}} & \cite{owens2021modeling} \\
    \hline
    \end{tabular} \label{numsim_param_table}
    \end{center}
    \caption{Baseline parameters.}
    \label{table:params}
    \end{table}
\end{flushleft}

Several studies have shown that CAR T-cells face several obstacles in penetrating solid tumors and reaching their intended targets within the tumor site \cite{martinez2019car}. Thus, we assume that (after converting to \unit{mm^3}), only 1\% of the CAR T-cells reaches the target tumor and set $I_C=0.1$ based on this discussion. At the parameter fitting stage, which is described in the subsequent paragraphs, initial conditions are set as $T_{s}(0)=50p $, $T_{r}(0)=50(1-p)$, $C(0)=0$ and $B(0)=0.1$ where $p$ is the percentage of the antigen-positive tumor cell population. All units are given in \unit{mm^3}. 

The control group associated with 100\% mesothelin positive case allows us to calibrate the growth rate $r_1$ of antigen-positive tumor cell population in Eq.\eqref{eq:Ts}. We then fix this value and calibrate the growth rate $r_2$ of antigen-negative tumor cell population using the control data associated with 90\% and 75\%  mesothelin positive cases. Through our analysis, we determined that $r_1 = 0.151$ and $r_2 = 0.180$. Our analysis indicates that, in this case, the tumor population lacking CART antigen expression exhibits a significantly higher growth rate compared to the population expressing the CART antigen, which also aligns with the experimental data plotted in Figure 1 of \cite{klampatsa2020analysis}.

\begin{figure}[H]
    \centering 
\begin{subfigure}{0.32\textwidth}
  \includegraphics[width=\linewidth]{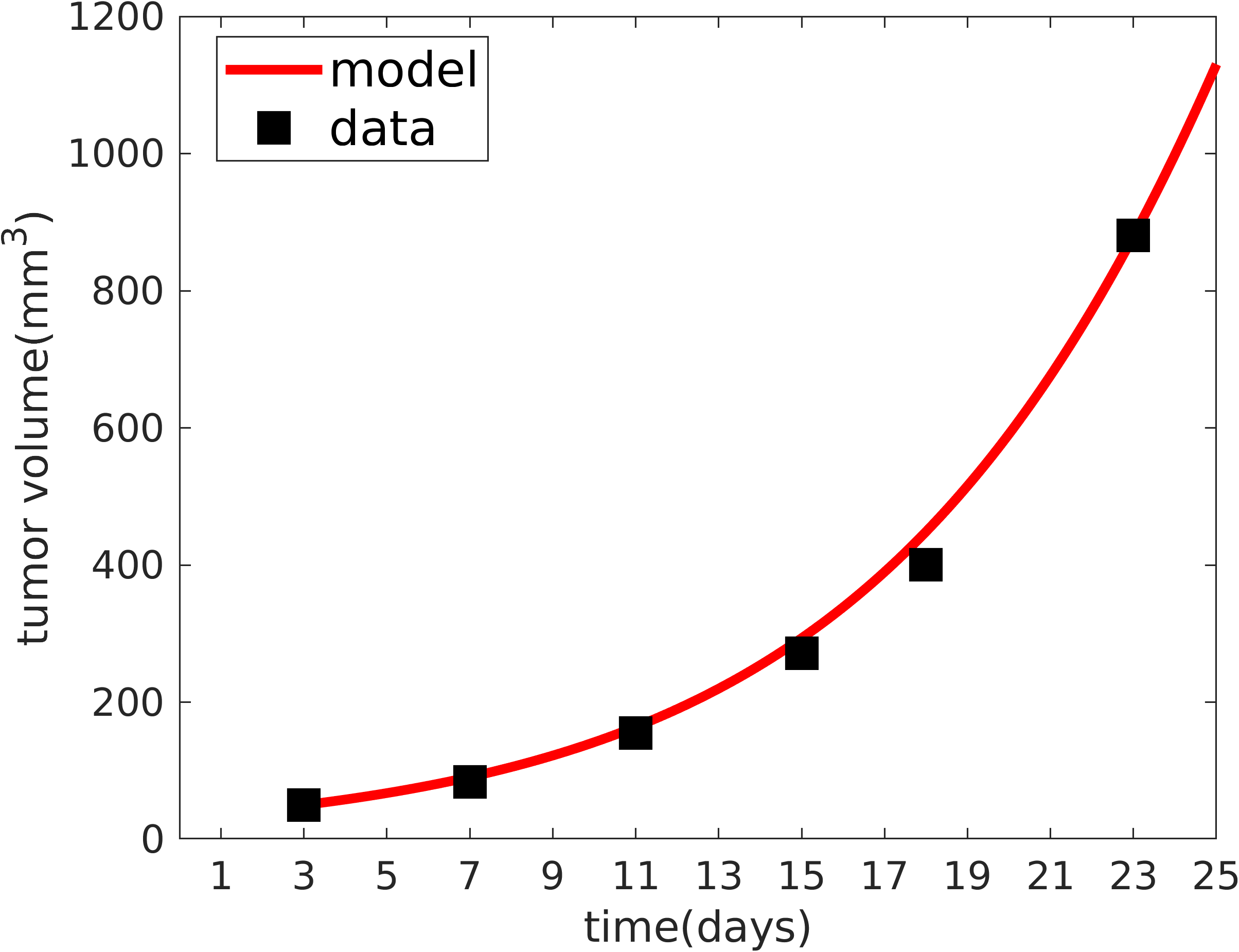}
\end{subfigure}\hspace{1mm} 
\begin{subfigure}{0.32\textwidth}
  \includegraphics[width=\linewidth]{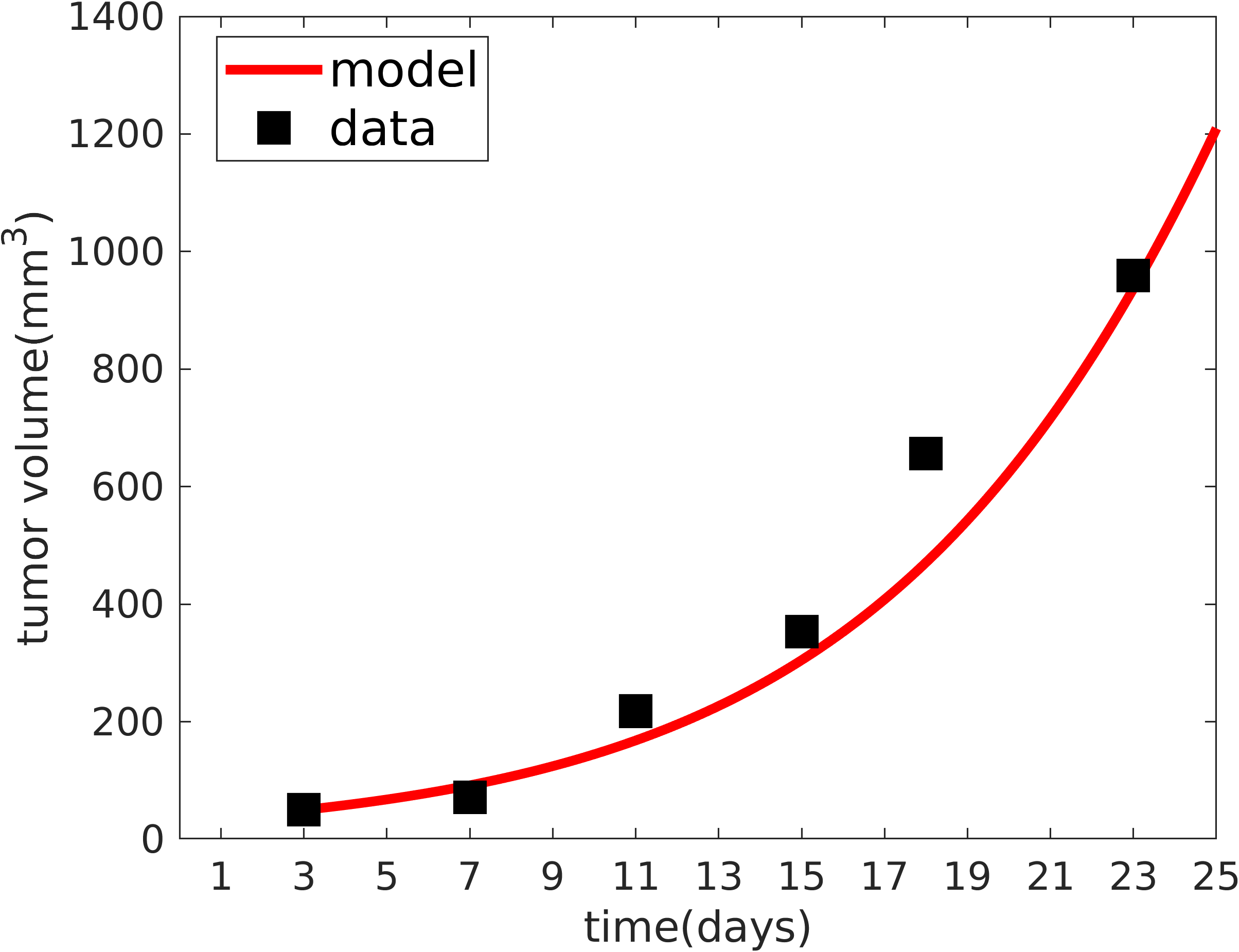}
\end{subfigure} 
\medskip
\begin{subfigure}{0.32\textwidth}
  \includegraphics[width=\linewidth]{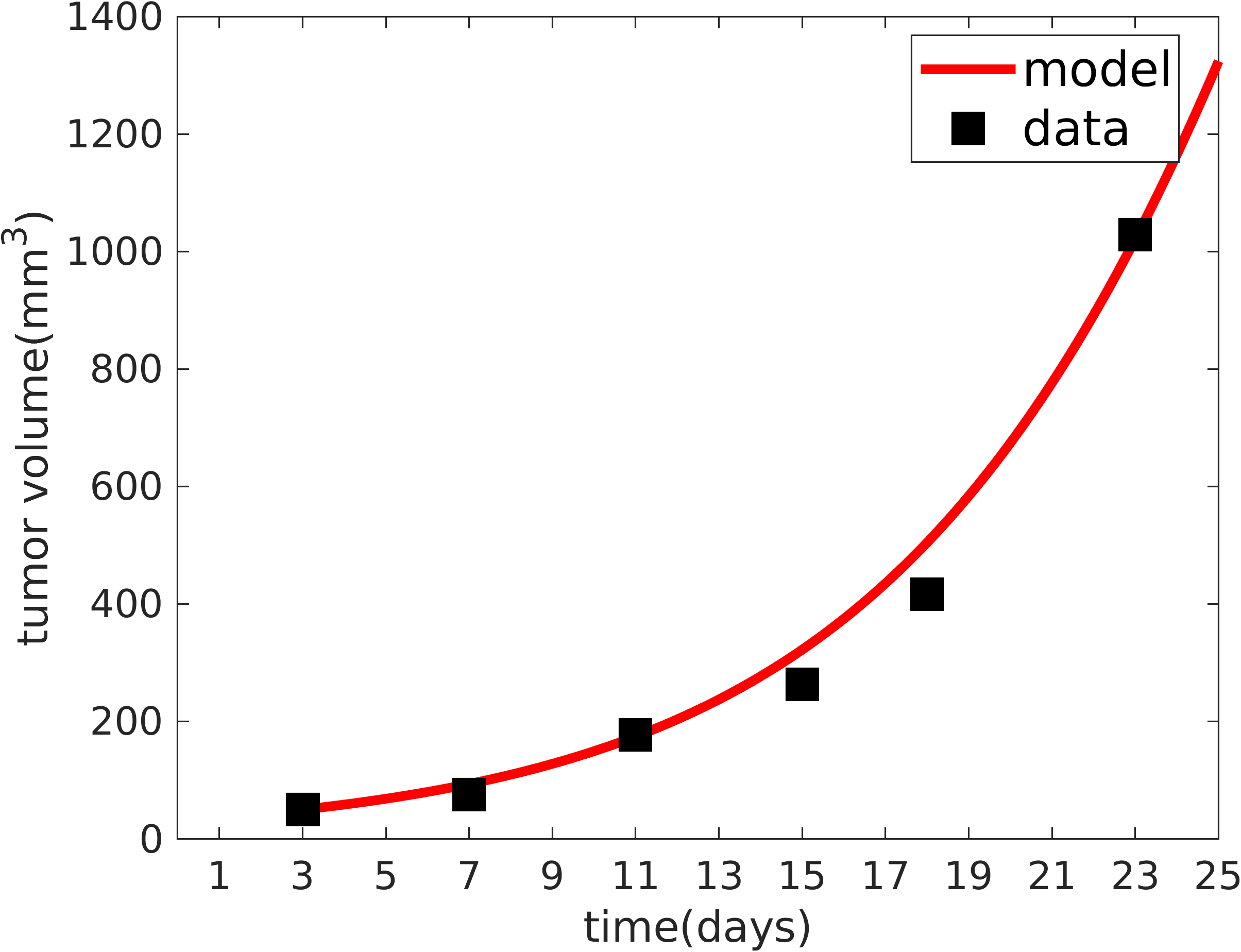}
\end{subfigure}
\caption{The image depicts a comparison between the model fit and the actual data in the control group, with the graph representing 100\%, 90\%, and 75\% antigen-positive cases on the left, middle, and right, respectively. The graph demonstrates a good agreement between the model fit and the observed data, indicating that the model accurately represents the behavior of the control group across varying levels of antigen positivity.}
\label{fig:control}
\end{figure}

The good agreement with the model and the data can be seen in Fig-\ref{fig:control}. Note that only first two ODE terms in Eq-\eqref{eq:Ts} and Eq-\eqref{eq:Tr} are active in case of control data since there are no treatment and bystander effect present.  

\begin{figure}[H]
    \centering 
\begin{subfigure}{0.32\textwidth}
  \includegraphics[width=\linewidth]{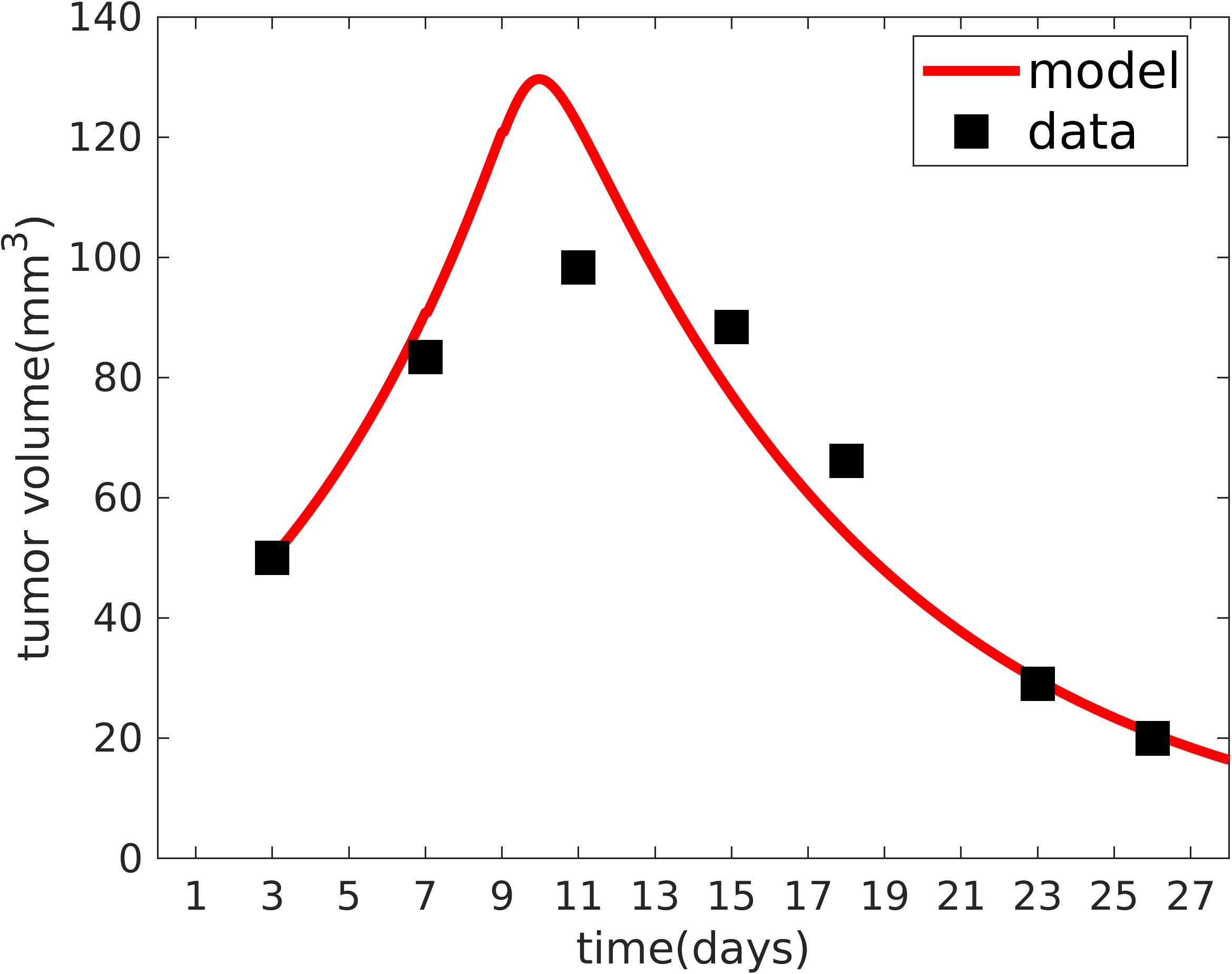}
\end{subfigure}\hspace{1mm} 
\begin{subfigure}{0.32\textwidth}
  \includegraphics[width=\linewidth]{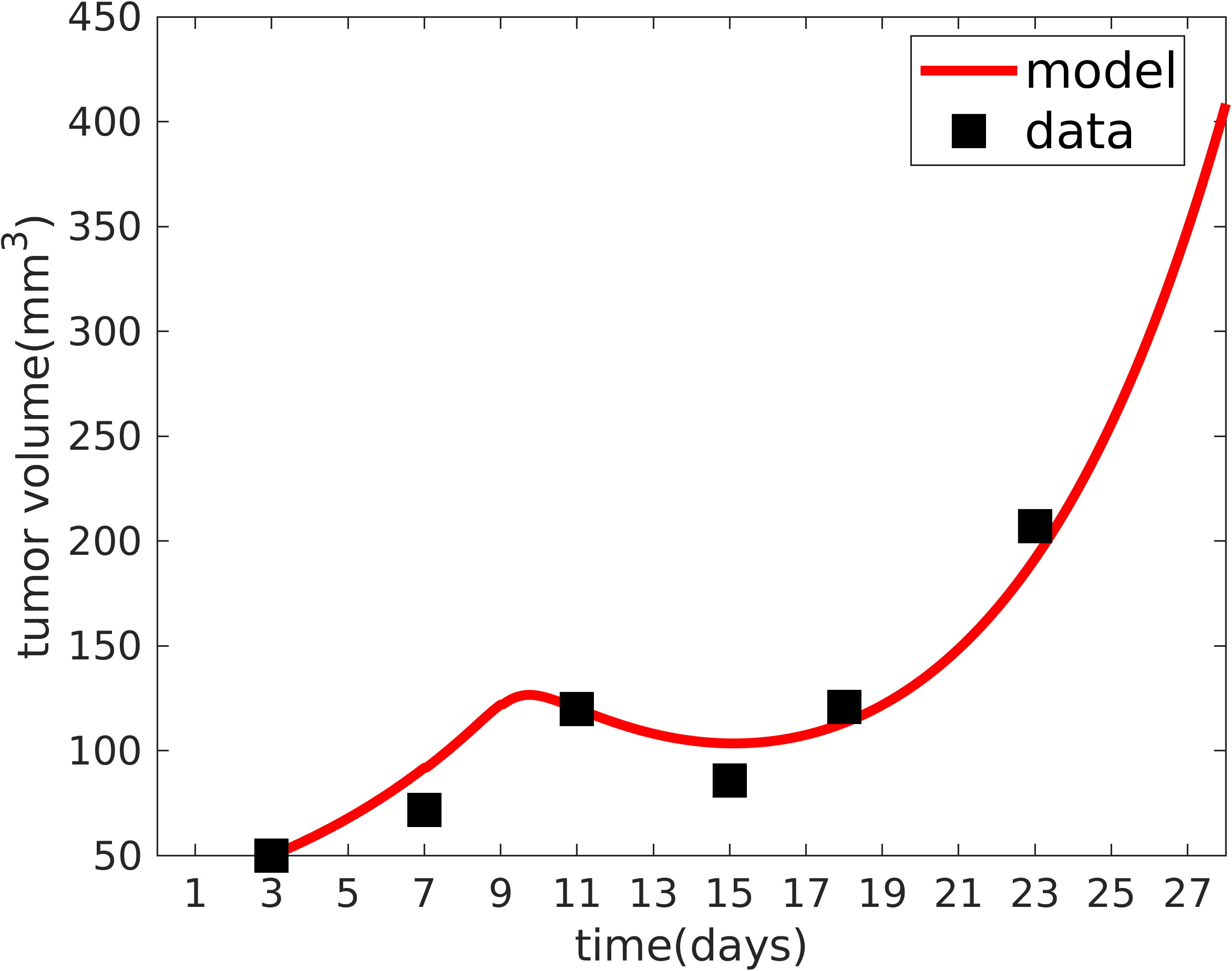}
\end{subfigure}
\begin{subfigure}{0.32\textwidth}
  \includegraphics[width=\linewidth]{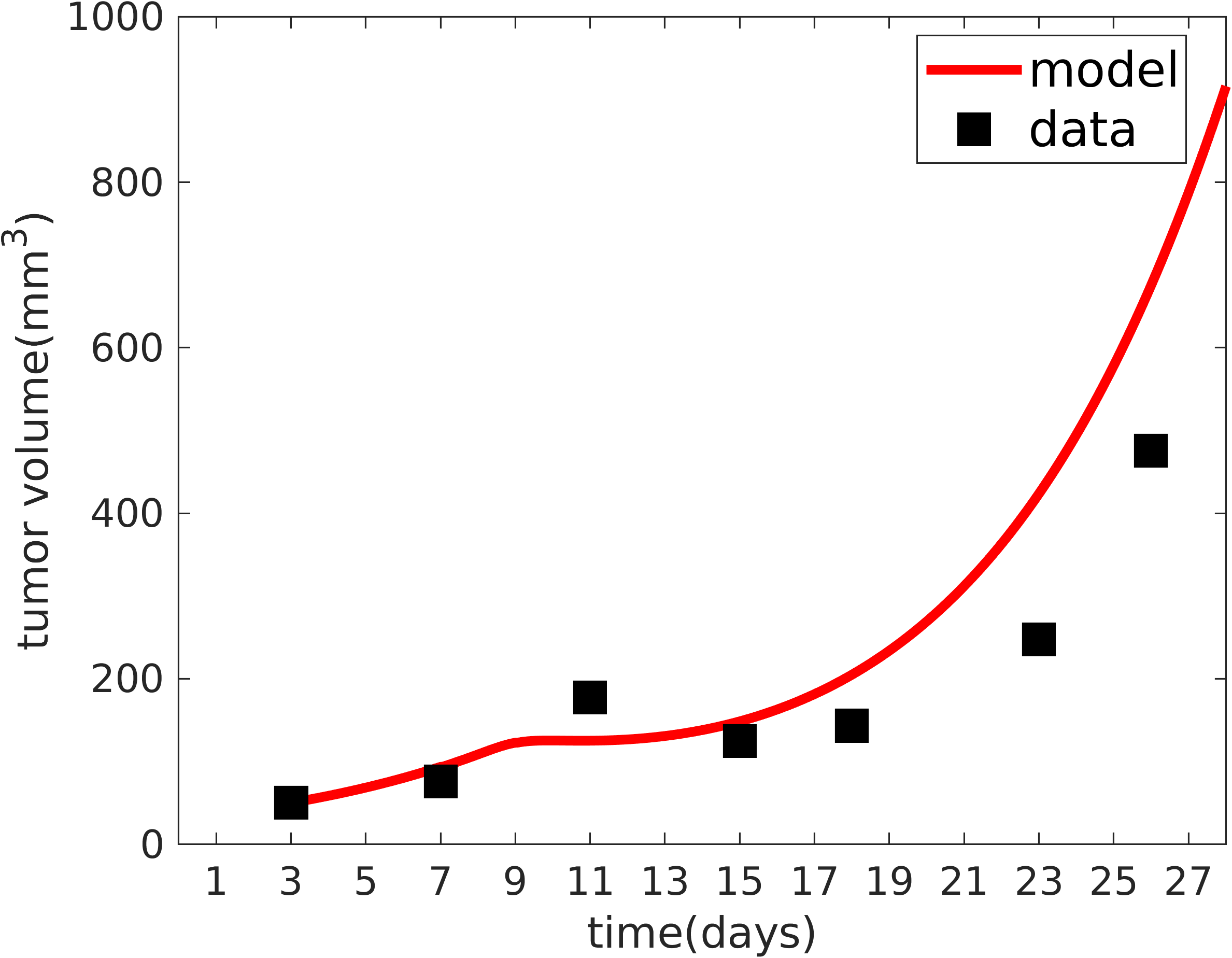}
\end{subfigure}
\caption{Comparison of model fit and data for CAR T-cell treatment. The figure on the left demonstrates the complete tumor elimination for 100\% antigen positive case. The figure in the middle and figure on the right are 90\% and 75\% antigen positive cells. We observe the absence of bystander effect in both cases. Although CAR T-cell treatment could not slowed tumor growth temporarily, it could not cure the tumor progression.
75\%(left), 90\%(middle) and 75\%(right) antigen positive cases. }
\label{fig:treatment}
\end{figure}

To calibrate the parameters associated with CAR T-cell treatment without bystander effect, we use the results of the two-dose CAR T-cell regimen. In this way, we calibrated the exponent of tumor lysis
$l$, maximum lysis rate of antigen-positive tumor cells due to CARTs $d_C$ and the maximum recruitment of CAR T-cells $\mu_C$. The results can be observed in Fig-\ref{fig:treatment}.

\begin{figure}[H]
    \centering 
\begin{subfigure}{0.40\textwidth}
  \includegraphics[width=\linewidth]{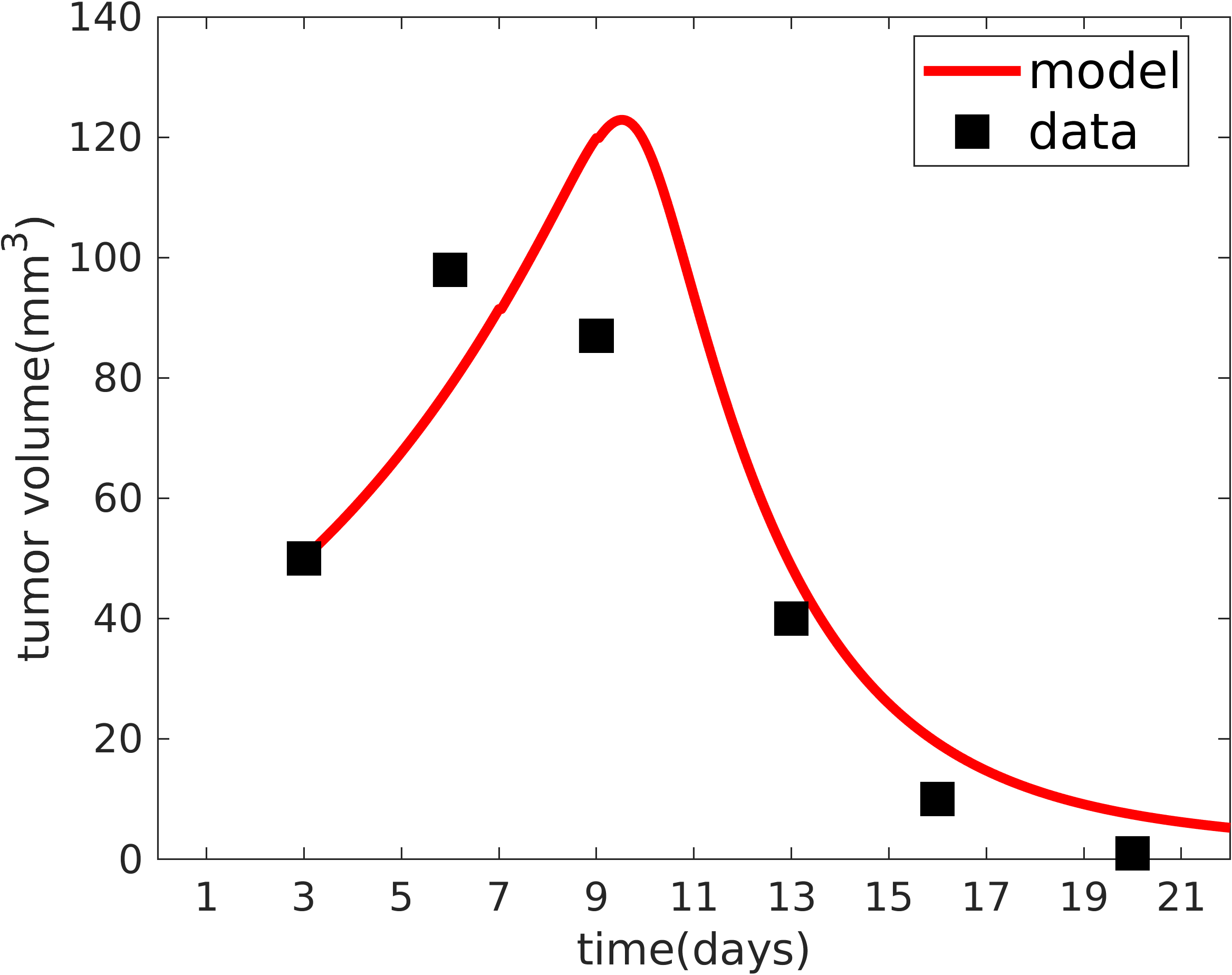}
\end{subfigure} 
\begin{subfigure}{0.40\textwidth}
  \includegraphics[width=\linewidth]{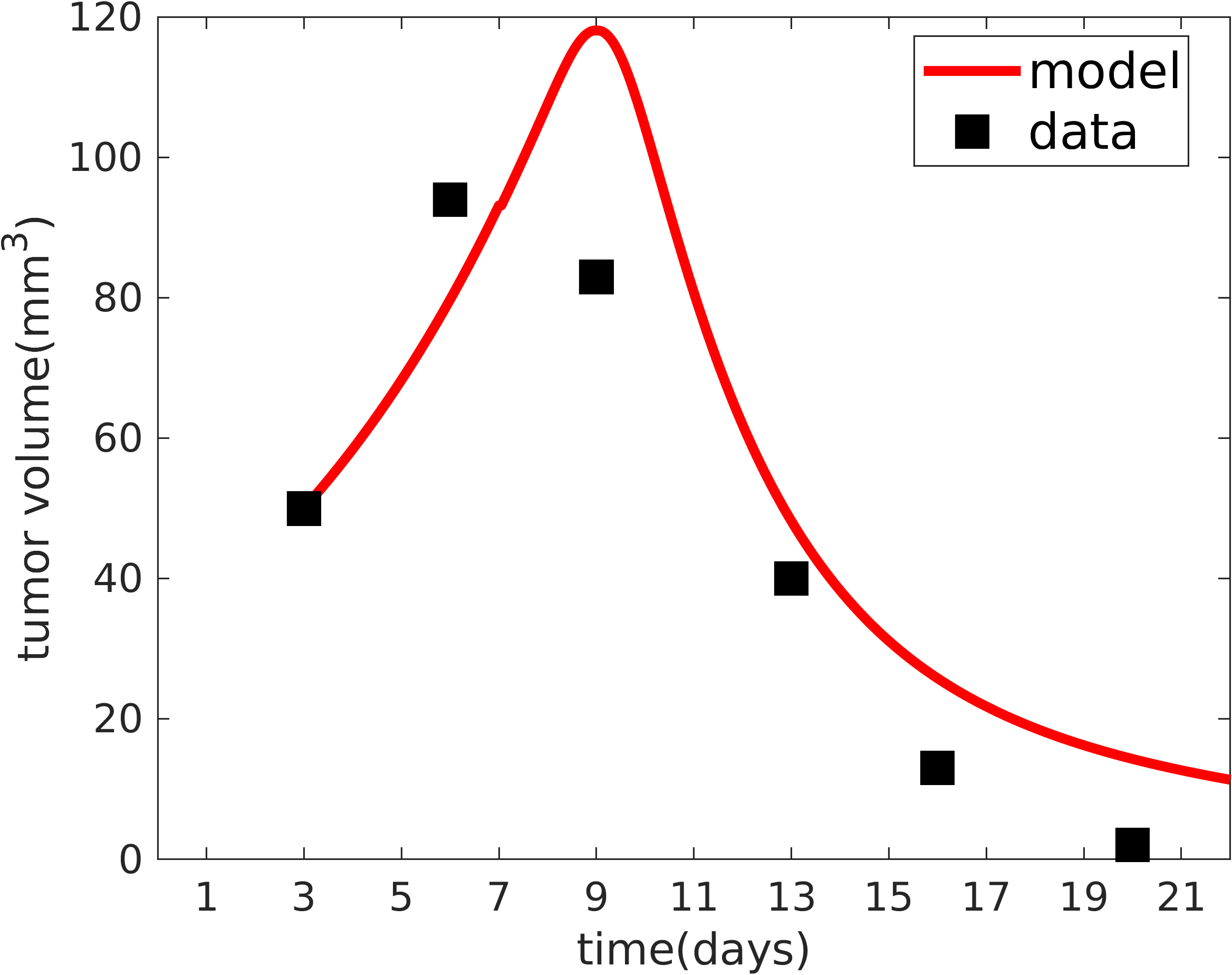}
\end{subfigure}
\caption{Comparison of model fit and data for CTX induced bystander effect on experiments conducted on 90\% and 75\% antigen positive tumor types. Administration of CTX along with M11 CAR T-cells cured the tumors in both cases proving the bystander effect.}
\label{fig:bystander}
\end{figure}

As mentioned above, CTX administered as a single dose(100\unit{mg/kg}) a day before the CAR T-cell treatment can induce bystander effect leading to tumor cure. This data allows us to calibrate maximum lysis rate of tumor cells due to bystander cells $d_B$ and the maximum recruitment of bystander  cells $\mu_B$. Fig-\ref{fig:bystander} displays these results.

\section{Results}
\label{sec:results}

\subsection{Tumor Response to Therapy is Largely Independent of CAR T-Cell Dose}

In Sec-\ref{sec:paramest}, we demonstrated that our model is good agreement with the experimental data. We will now simulate different scenarios to gain a better understanding about bystander effect of CAR T-cells. 

\begin{figure}[H]
    \centering 
\begin{subfigure}{0.85\textwidth}
  \includegraphics[width=\linewidth]{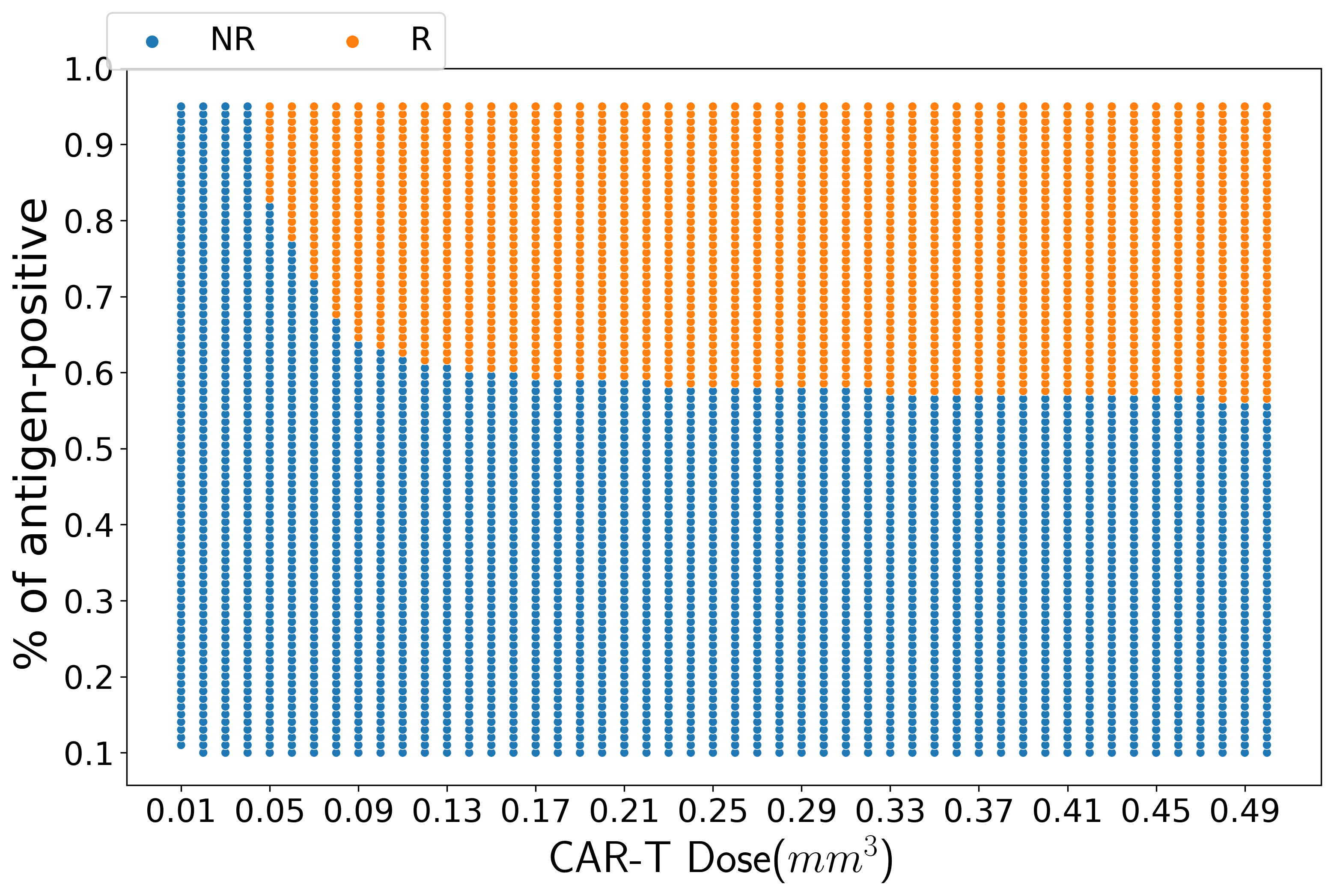}
\end{subfigure} 
\caption{Therapeutic responses for varying cart cell dosages and antigen positive percentages. We can observe that increasing the dosage of CAR T-cells does not guarantee a favorable therapeutic outcome. Specifically, in cases where the levels of CAR T-cell antigen expression are approximately 55\% or lower, augmenting the quantity of injected CAR T-cells does not result in a responsive treatment. We can also observe that the minimum dose required to elicit a response is around 0.05 \unit{mm^3}, which is below the baseline dose 0.1 \unit{mm^3} we utilized in the model.}
\label{fig:scatter}
\end{figure}

In Figure-\ref{fig:scatter}, we present the results of a study evaluating the therapeutic responses to a range of CAR T-cell dosages and levels of antigen positivity over a period of 45 days. Our analysis involved classifying the response of each tumor based on whether the final tumor burden was less than the initial seed size of the solid tumor (2 \unit{mm^3}) prescribed in \cite{klampatsa2020analysis}. If the final tumor population is below this threshold, we label it  as a responder(R),  denoted by orange markers, otherwise we label the tumor as a non-responder(NR) and denoted by blue markers. Our findings indicate that simply increasing the dosage of CAR T-cells does not guarantee a favorable therapeutic outcome. Specifically, in cases where the levels of CAR T-cell antigen expression are approximately 55\% or lower, augmenting the quantity of injected CAR T-cells does not result in a responsive treatment. 

\subsection{Virtual Patient Analysis Predicts Tumor Response Patterns}
\label{sec:VPs}

In this section, we generate a virtual patient (VP) cohort to ensure the exclusion of that reflects the variability in the data to examine the impact of the variability on treatment outcomes. Utilizing virtual patients allow us to conduct statistical analysis and draw conclusions on the effectiveness of various treatments or treatment combinations. We apply the treatment regimen described above to each VP in the cohort and identify any patterns or trends that may contribute to successful therapeutic outcome.

\begin{figure}[H]
    \centering 
\begin{subfigure}{0.32\textwidth}
  \includegraphics[width=\linewidth]{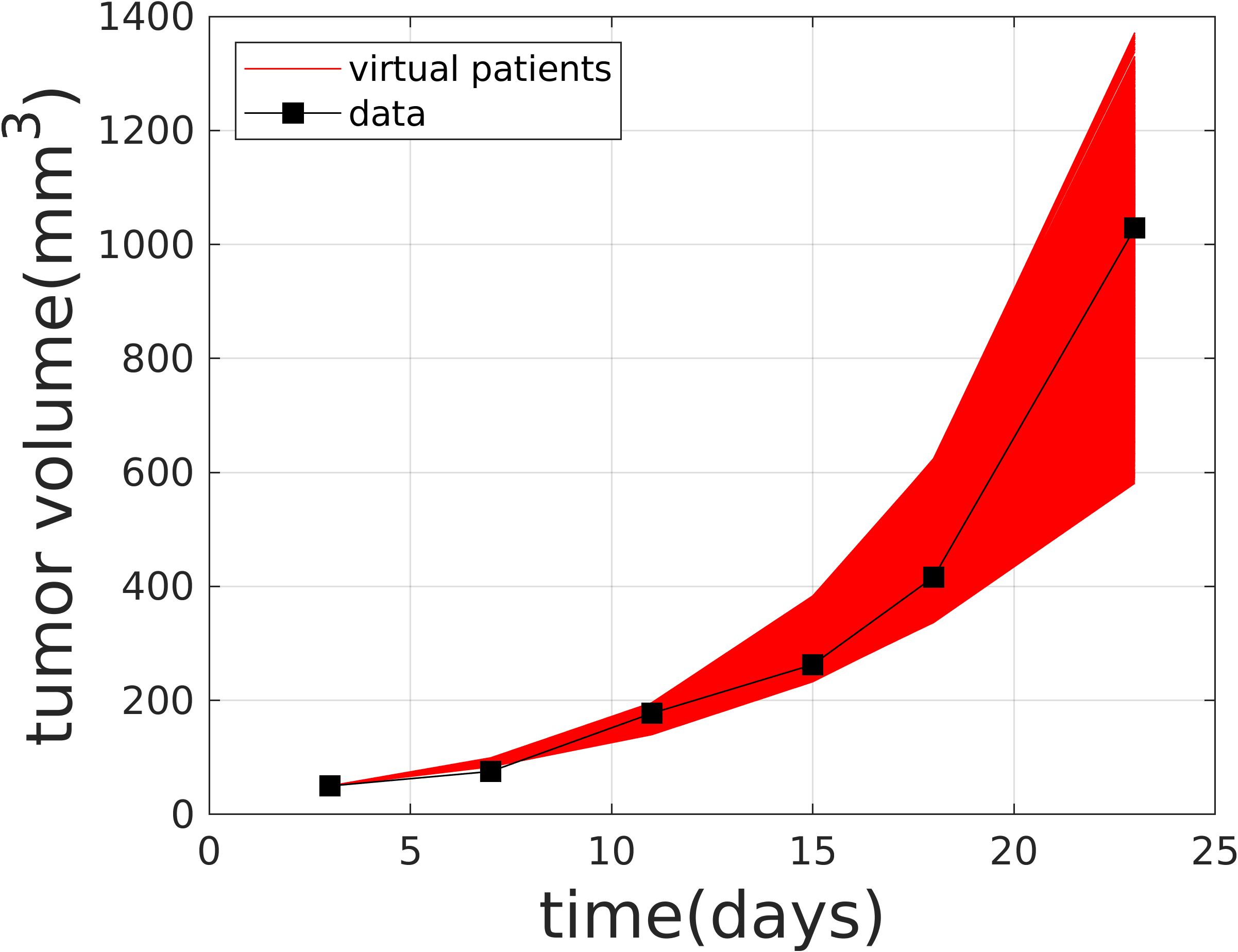}
\end{subfigure}\hspace{1mm} 
\begin{subfigure}{0.32\textwidth}
  \includegraphics[width=\linewidth]{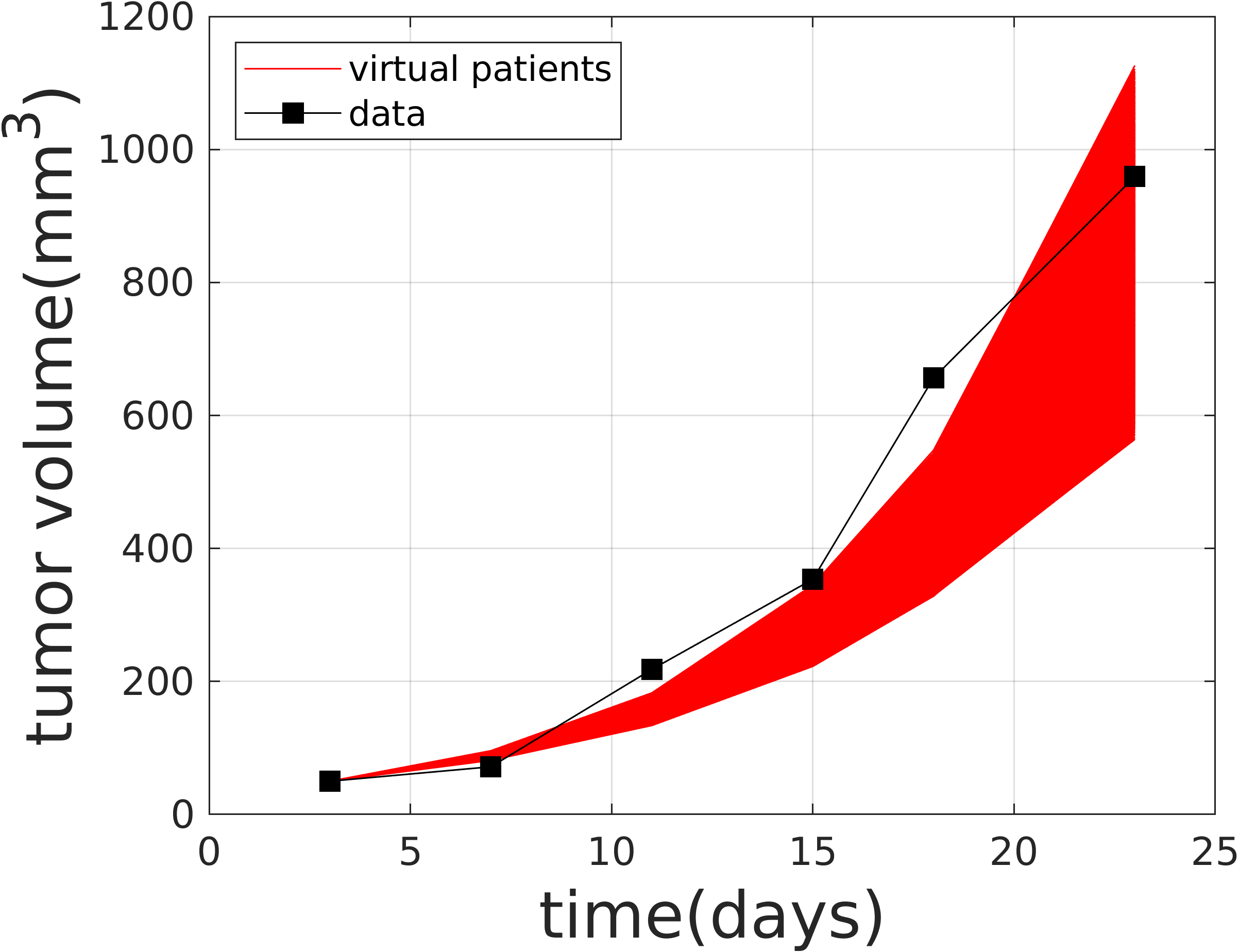}
\end{subfigure}
\begin{subfigure}{0.32\textwidth}
  \includegraphics[width=\linewidth]{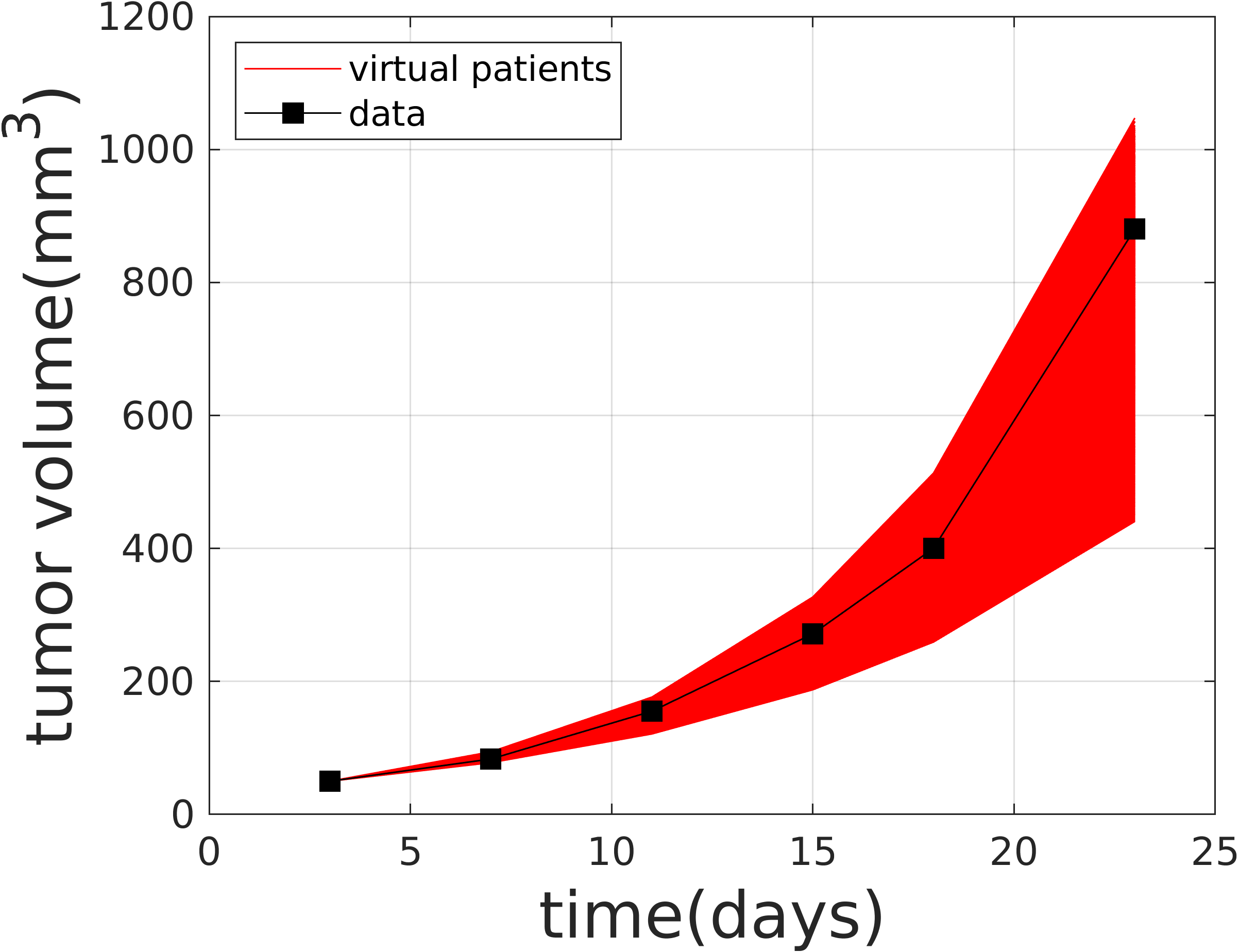}
\end{subfigure}
\caption{Comparison of VPs and the control data. Trajectories of 6433 control VPs are illustrated in region while the control data regarding 75\%,90\% and 100\% antigen-positive tumor population are represented with black marker. We can observe that the generated VPs cohort is a good representative of the actual experimental data.}
\label{fig:controlVP}
\end{figure}

To generate virtual patients (VPs), we begin by considering the parameters $r_1$, $r_2$, and $K_1$, and generate values for each parameter such that they fall within a 0.5 relative error band of each data point with respect to the control data for the 100\%, 90\%, and 75\% cases. By imposing this acceptance criteria we aim to obtain a set of parameters, $r_1$, $r_2$, and $K_1$, that capture the dynamics in the experimental data. In total, we generated \textbf{6433} such values using this methodology, as illustrated in Figure-\ref{fig:controlVP}. Parameters of interests fall in the following ranges; $r1 \in [0.11, 0.16], r2 \in [0.14,0.22], K1 \in [2\times 10^3,7\times 10^3]$.

In order to establish a VP cohort that encompasses all 16 variables, we fix the values of $r_1, r_2,K_1$ obtained above, and for each of the control VP, we generate the remaining 13 parameters by sampling from a uniform distribution within the following parameter ranges.

\vspace{-30pt}

\[
\begin{aligned}
l &\in [0.5, 1.8], \quad d_C \in [0,1], \quad \mu_C \in [0,1], \quad s \in [10^{-1}, 5 \times 10^{-1}],
\quad \gamma_C \in [10^{-2}, 5 \times 10^{-2}], \\
\quad k &\in [10^{-7}, 3 \times 10^{-7}], \quad w_C \in [10^{-5}, 5 \times 10^{-5}], \quad
K_2 \in [2 \times 10^{3}, 7 \times 10^{3}], \quad d_B \in [0,1], \quad \mu_B \in [0,1], \\
b &\in [10^{-2}, 10^{-1}], \quad \gamma_B \in [10^{-2}, 10^{-1}], \quad w_B \in [10^{-6}, 10^{-5}].
\end{aligned}
\]

\begin{figure}[!ht]
    \centering 
    \begin{subfigure}{0.75\textwidth}
  \includegraphics[width=\linewidth]{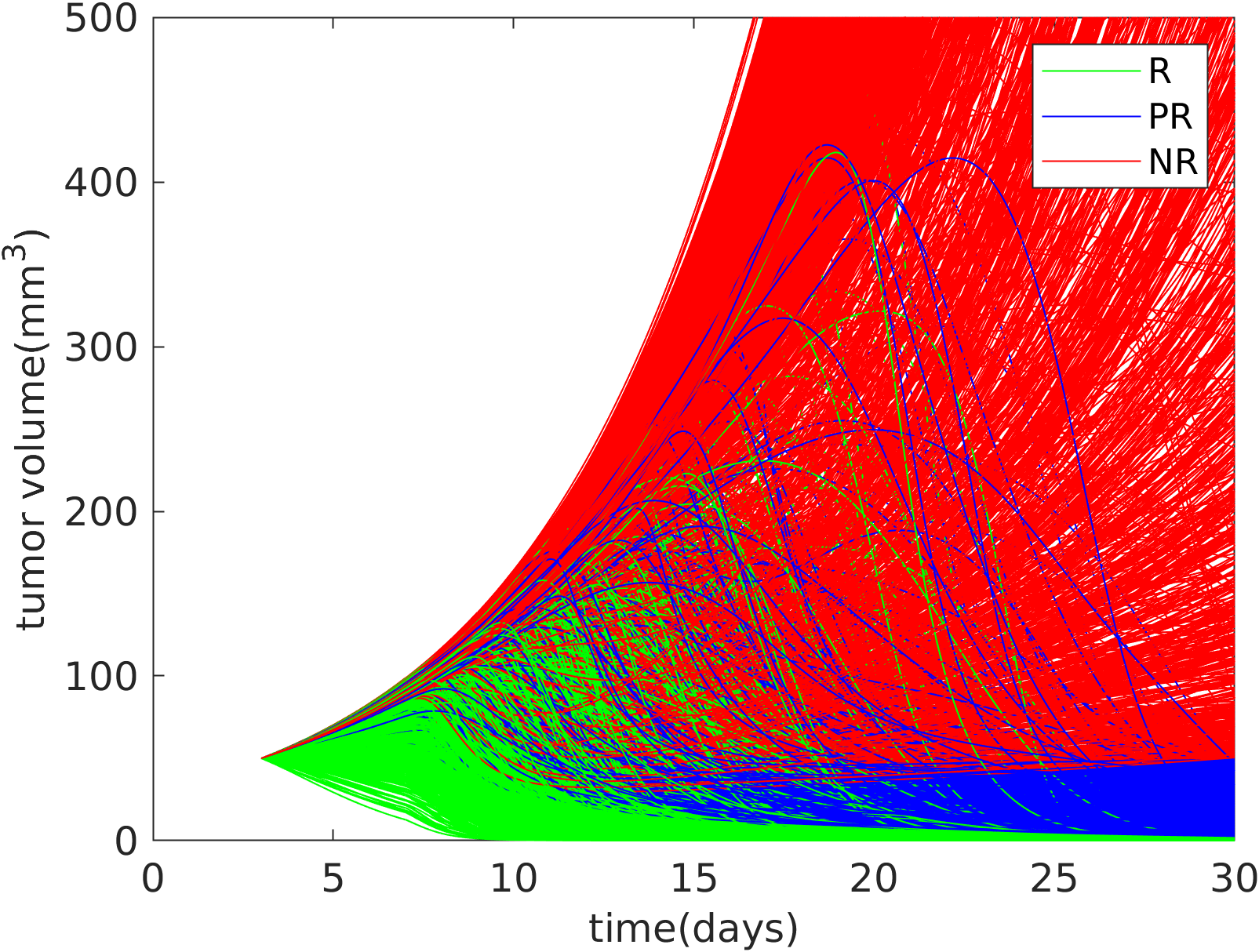}
\end{subfigure}
\caption{Trajectories of 6433 VPs are illustrated based on the final therapeutic outcome. Blue, green and magenta trajectories represent R,PR and NR,respectively. Out of
6433 VPs, the response distribution in terms of percentage occurs as 53\% R, 11 PR and 36\% NR.}
\label{fig:base_response}
\end{figure}

We will then submit these VPs to standard treatment regimen and investigate which patients responds to the treatment. We also fix the antigen positive percentage to $75\%$. As previously mentioned, the initial seed size of the solid tumor for all experiments in \cite{klampatsa2020analysis} was 2 \unit{mm^3}, and the CAR T-cell treatment was initiated when the tumor size reached 50 \unit{mm^3}. Thus, we will use the following criteria to classify the outcome of the treatment for each VP. 

\[ response=\begin{cases} 
      R, & T_{45}\leq 2 \unit{mm^3} \\
      PR, & 2 \unit{mm^3}  \leq T_{45} \leq 50 \unit{mm^3}  \\
      NR, & T_{45} \geq 50 \unit{mm^3}
   \end{cases}
\]

where R, PR and NR represent response, partial-response and no-response, respectively. We display the trajectory of each VP in Fig-\ref{fig:base_response} where blue, green and magenta trajectories represent R, PR and NR categories, respectively. $T_{45}$ is the size of the final tumor population at day $t=45$, given in $\unit{mm^3}$. Out of 6433 VPs, the response distribution in terms of percentage occurs as follows
\begin{equation*}
    \mbox{R}:53\%,\mbox{PR}:11\%,\mbox{NR}:36\%
\end{equation*}

\newpage
\subsection{Enhancing Bystander Cell Cytotoxicity Increases Survival Rate}
\begin{figure}[!ht]
    \centering 
\begin{subfigure}{0.95\textwidth}
  \includegraphics[width=\linewidth]{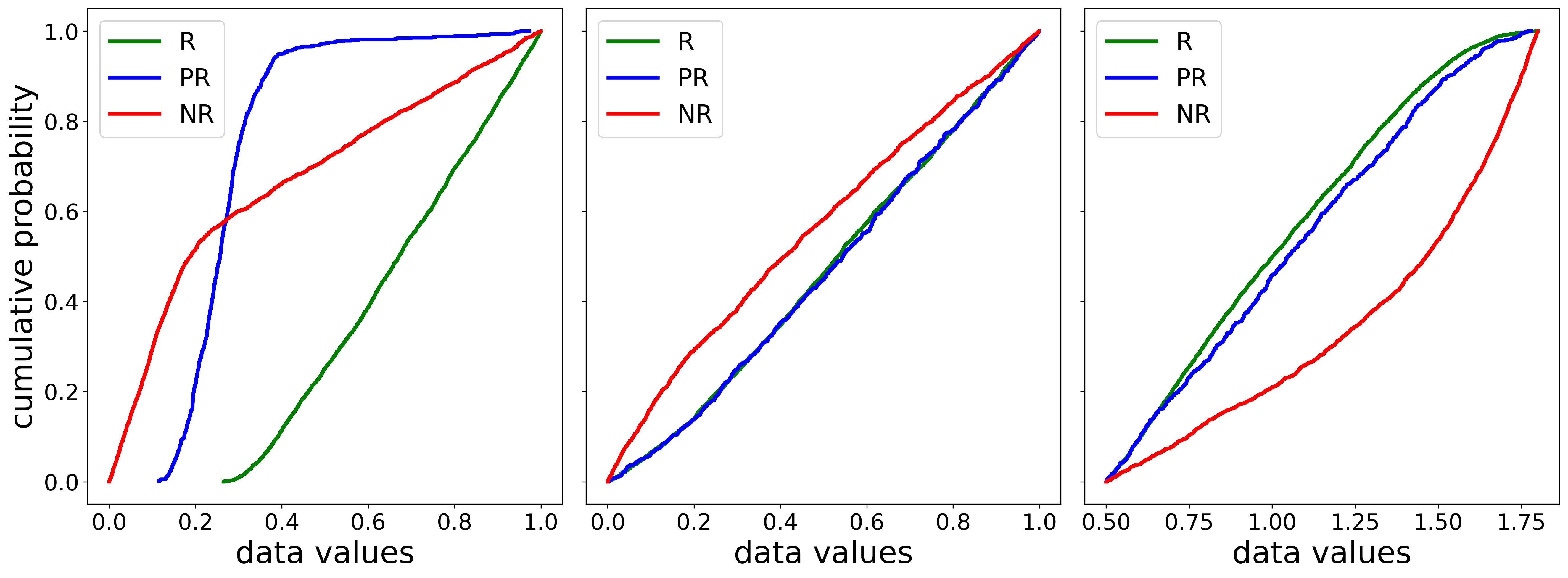}
\end{subfigure}
\medskip
\caption{Empirical cumulative distribution plots for $d_B$(left), $d_C$(middle) and $l$(right). We observe that the distributions of these parameters are significantly different across the three classes, R, PR, and NR. For example, the left figure associated with $d_B$ shows that responders are strongly associated with higher $d_B$ values while this case is the opposite for non-responders.}
\label{fig:ecd}
\end{figure}

We can now assess the statistical significance between these three groups and determine which parameters have a greater impact on the treatment outcome. One way of achieving this task would be to inspect the empirical cumulative distribution function(ECDF) of each parameter across these three classes. ECDF is constructed by ordering the data points from smallest to largest and plotting the proportion of data points that are less than or equal to each value. The resulting curve shows how the data is distributed, with a steeper curve indicating a higher concentration of data points at lower values. We use ECDF because it is non-parametric, meaning the method does not make any assumptions about the underlying distribution of the data, which matches our VP set-up. Through empirical cumulative distribution function (ECDF), we observe that the distributions of $d_B$, $d_C$, and $l$ are significantly different across the three classes, R, PR, and NR, as seen in Fig-\ref{fig:ecd}. For example, the left figure associated with $d_B$, maximum killing rate of bystander cells, shows that responders are strongly associated with higher $d_B$ values while this case is the opposite for non-responders.

To gain a deeper understanding, we can also visualize box plots that show the distribution of values of the corresponding parameter within each class. The box plots reveal that the distribution of $d_B$ values is noticeably different across the three classes. Specifically, the box plot for the responder class shows a higher median and a larger spread of values compared to the other two classes, indicating that the $d_B$ values tend to accumulate towards the upper range for this class. In contrast, the box plot for the non-responder class shows a lower median and a smaller spread of values, indicating that $d_B$ values for this class tend to be lower and are more tightly clustered towards the lower range. These differences suggest that $d_B$ may be a key parameter in distinguishing between the R and NR classes. Similarly, we observe that the $l$ values for the non-responders  tend to be higher compared to the other two classes.


\begin{figure}[!ht]
    \centering 
\begin{subfigure}[t]{0.45\textwidth}
  \centering
  \includegraphics[height=2in]{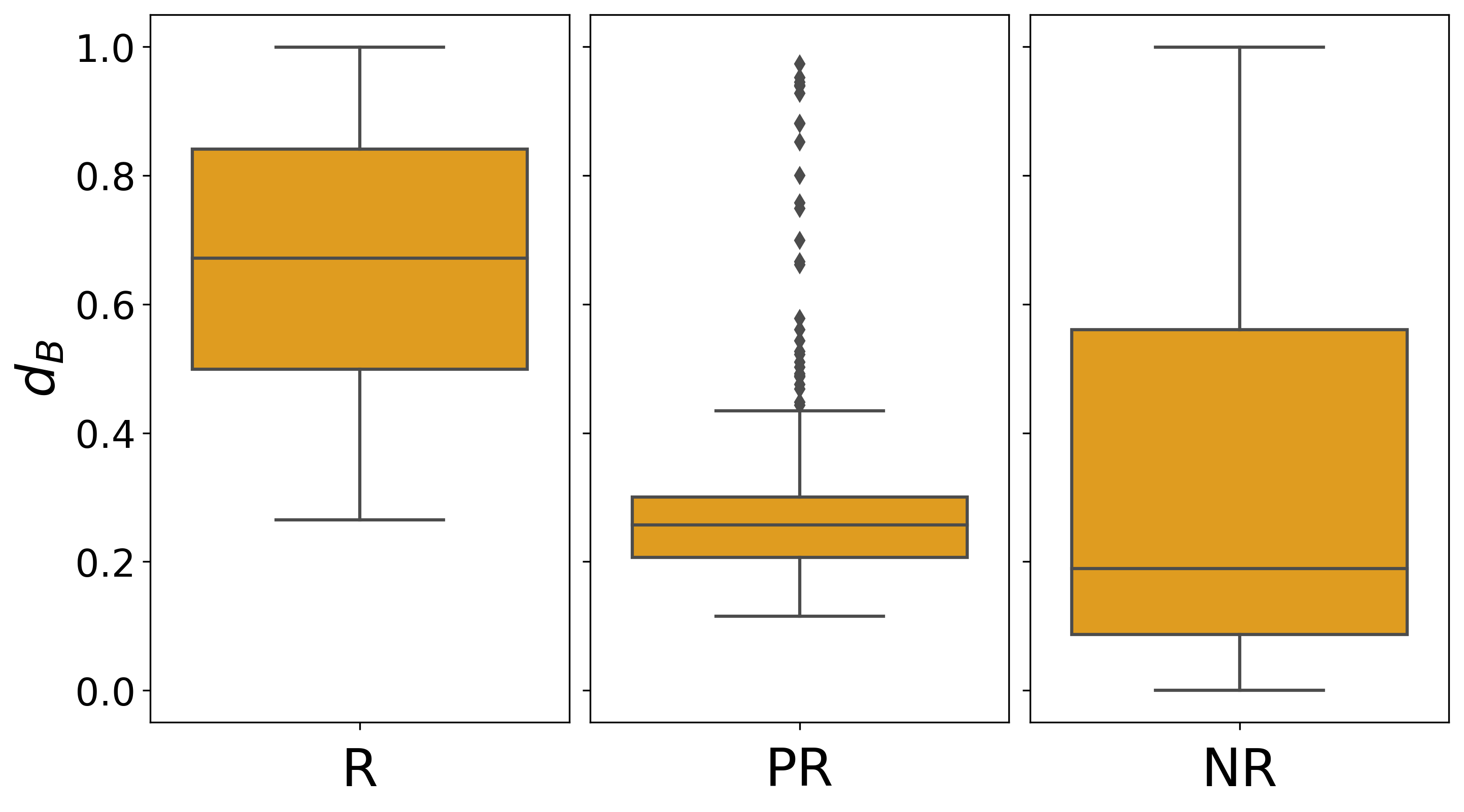}
\end{subfigure}%
\hspace{30pt}
\begin{subfigure}[t]{0.45\textwidth}
  \centering
  \includegraphics[height=2in]{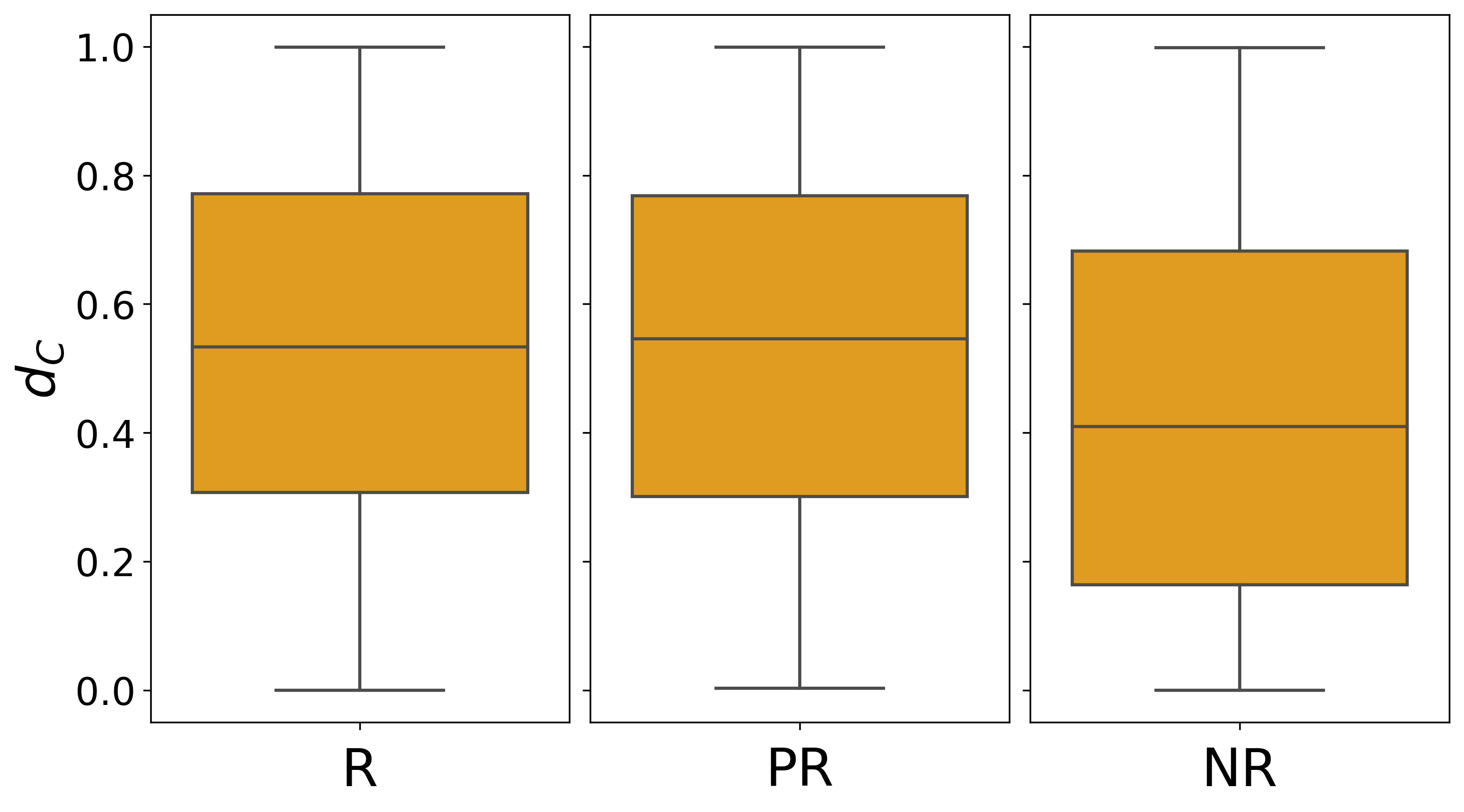}
\end{subfigure}
\medskip
\begin{subfigure}[t]{0.45\textwidth}
  \centering
  \includegraphics[height=2in]{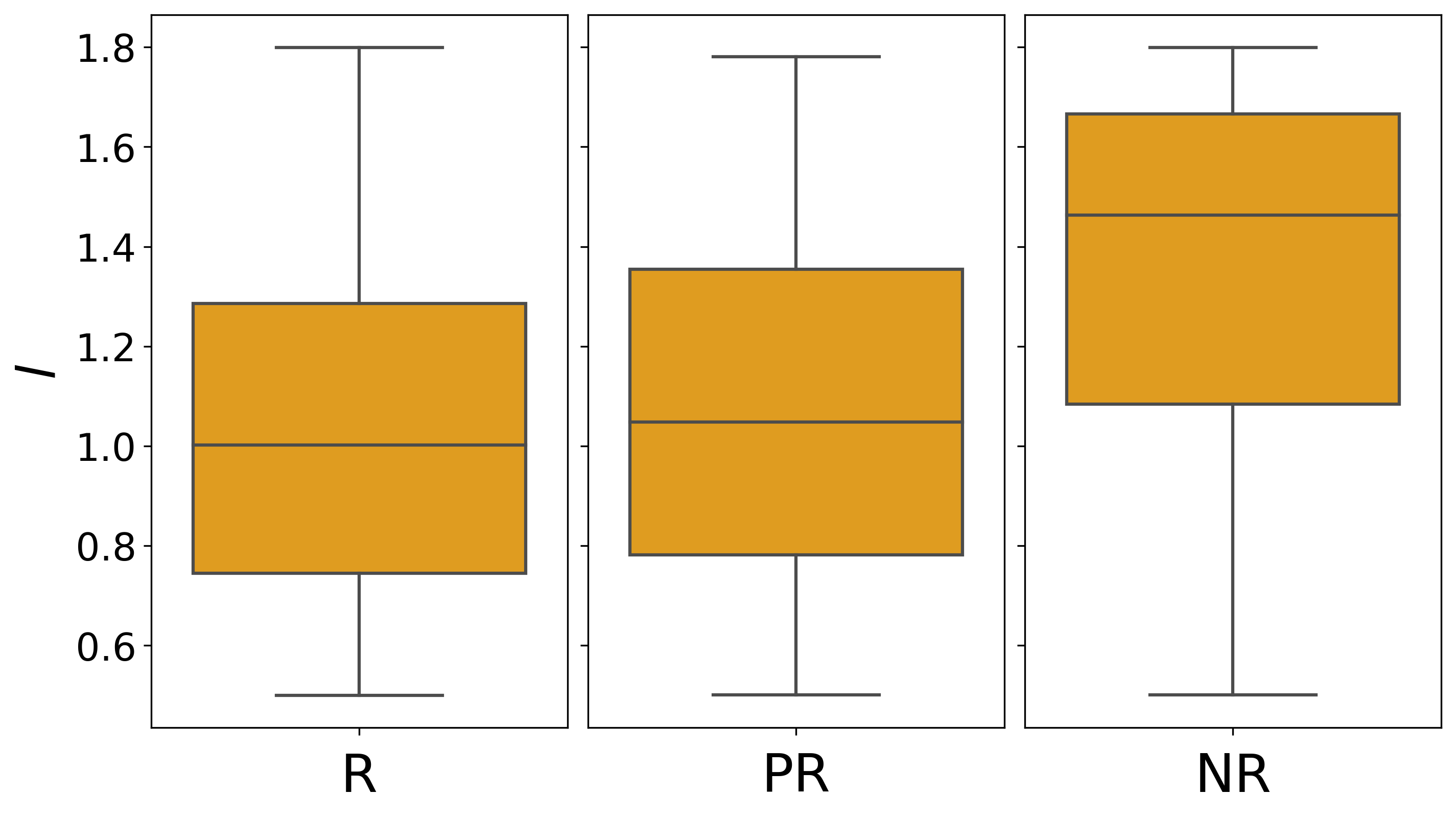}
\end{subfigure}
\caption{Distribution profile of $d_B$(bottom left), $l$(bottom right) and $d_C$(middle) for the categories of interests. Notice that the box plot for the responder class regarding $d_B$ shows a higher median and a larger spread of values compared to the other two classes, indicating that responders are associated with higher $d_B$ values. This suggest that higher maximum killing rates may result in an improved therapeutic outcome.}
\label{fig:boxdB}
\end{figure}

\begin{figure}[H]
    \centering 
\begin{subfigure}{0.85\textwidth}
  \includegraphics[width=\linewidth]{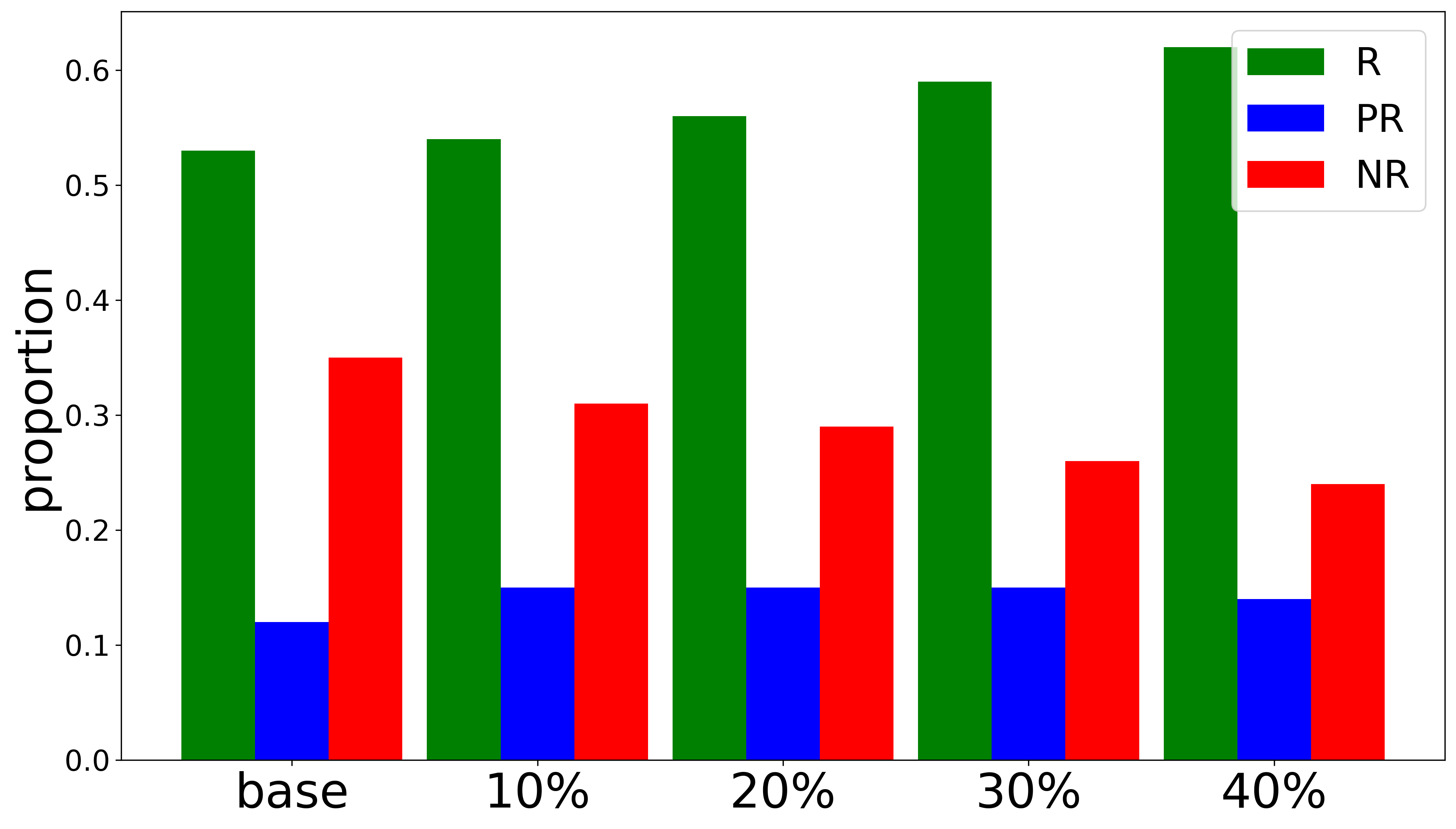}
\end{subfigure} 
\caption{Each incremental increase in $d_B$ results in a higher proportion of responders compared to the baseline scenario. This suggests that  enhancing the cytotoxic capacity of bystander cells is is a crucial aspect to maximize the bystander effect.}
\label{fig:dB_bar}
\end{figure}

Recall that $d_B$ represents the maximum killing rate of bystander cells once they are activated. Through our VP cohort, we can investigate whether a higher maximum killing rate would result in an improved therapeutic outcome. To achieve this, we systematically increased the value of $d_B$ by varying percentages while keeping all other parameters constant. We then submitted each virtual patient to the same standard therapy described above. As depicted in Figure-\ref{fig:dB_bar}, each incremental increase in $d_B$ resulted in a higher proportion of responders compared to the baseline scenario.  Based on this observation, our model suggests that maximizing the bystander effect may be a more effective approach to achieving successful outcomes in CAR T-cell therapy compared to solely increasing the number of CAR T-cells.

\section{Discussion}
\label{sec:discuss}

In this study, we developed a mathematical model to investigate the bystander effect in CAR T-cell therapy for solid tumors. Our model exhibits a remarkably accurate fit to the unconventional data obtained from an in-vivo experiment. To the best of our knowledge, this is the first mathematical investigation specifically focused on the bystander effect in CAR T-cell treatment. The findings of our study are consistent with key observations made by \cite{klampatsa2020analysis}. In their study, the authors observed that the bystander effect was not influenced by the augmented persistence of administered CAR T-cells. Instead, they provided evidence demonstrating that the presence of endogenous CD8 T cells was a necessary factor for the occurrence of the bystander effect. Our results also indicate that augmenting the number of CAR T-cells reaching the tumor site does not enhance the therapeutic outcome, even in cases of high antigen expressivity, suggesting the absence of bystander effect. However, our model further reveals that enhancing the cytotoxic capability of bystander cells yield significantly improved outcomes, as revealed through virtual patient analysis.

It is important to highlight that our model incorporates components from well-established works cited in the model development section. In particular, our modeling approach aligns with a similar philosophy with \cite{owens2021modeling}, which explores the impact of CAR T-cell therapy in the presence of various chemotherapy regimens, primarily for hematologic malignancies. The key variables in this context are tumor cells, CAR T-cells, effector T-cells, and chemotherapy drugs. In our work, we break down the tumor population based on whether they can be targeted by CAR T-cells. Moreover, our model is primarily applicable to solid tumors since a mixing model is feasible only for solid tumors. Here, mixing refers to the experiment in which a mixture of tumor cells with different antigen expression profiles is used. Also, we do not specifically model drug infusion in our model due to the the nature of bystander experiment discussed in \cite{klampatsa2020analysis}. The authors therein did not administer a high dose of CTX since the goal was to investigate the effect of low-dose CTX on bystander effect, and determine if this dose could enhance the anti-tumor immune response without causing anti-tumor effects or lymphodepletion which essentially invalidates the observablity of bystander effect.

We would like to emphasize  that our model is driven by the experimentally available data. Therefore, one natural limitation of the current model is that we can investigate the bystander phenomena primarily from the perspective of CAR T-cells and bystander CD8 T-cells. However, given the good agreement with the experimental data, our model offers a foundation for further improvements and provides valuable insights into existing theories or the development of new hypotheses regarding this phenomenon. It is important to note that the study of bystander phenomena remains an active research area with ongoing advancements. Building upon the existing research, below we outline several potential research questions that can be explored within the same modeling framework. Interested readers can refer to \cite{klampatsa2020analysis} for a more comprehensive discussion.

It is clear that the bystander effect requires the presence of CD8 positive T cells. However, it is not clear exactly by which mechanism those effector T cells are activated. In this sense, one can investigate the role of CD4 helper T-cells or regulatory T cells(Tregs). In the context of the data we use, the latter is of special importance. Tregs are a specialized subset of T cells that play a crucial role in immune regulation and maintaining immune tolerance. In the context of cancer immunotherapy, Tregs can have a suppressive effect on the immune response against tumors \cite{whiteside2015role}. This suppression occurs through various mechanisms, including the inhibition of anti-tumor responses by endogenous CD8 T cells. On the other hand, CTX has been shown to have the ability to reduce the population or function of Tregs \cite{nakahara2010cyclophosphamide,humphries2010role}. By reducing the number of Tregs, CTX can potentially alleviate their suppressive effects on immune responses, particularly those directed against tumors. This pathway can be explored via mathematical modelling by incorporating Tregs and CTX into the existing framework.

Cytokines produced by CAR T-cells are another direction that can be explored. Upon activation, CAR T-cells can release cytokines such as IFN-gamma (IFNg) and tumor necrosis factor-alpha (TNF-a) \cite{silveira2022cytokines,cosenza2021cytokine}. These cytokines have the potential to directly kill tumor cells. Additionally, the released cytokines can activate the components of innate immune system, including macrophages, neutrophils, or natural killer (NK) cells. Once activated, these immune cells can induce tumor destruction through mechanisms such as phagocytosis. Although a fully coupled modelling scenario is possible,  the terms related to cytokines and innate immune system can be potentially decoupled from the other model variables and directly linked to CAR-T dynamics since this pathway is primarily triggered by CAR-T activation.

Exploring the concept of \textit{cross-presentation}, also known as \textit{T-cell cross priming}, could also be a promising line of investigation. This refers to the process by which antigen-presenting cells, particularly dendritic cells (DCs), present exogenous antigens(antigens that originate from outside the body) to CD8 T cells \cite{sanchez2017antigen}. DCs are immune cells that are crucial for orchestrating immune responses. They have a pivotal role in determining whether CD8 T cells, which are important for anti-tumor immune responses, will mount an immune response against tumor antigens or exhibit tolerance. DCs can present antigens from engulfed or captured tumor cells to CD8 T cells leading to the activation of CD8 T cells, enabling them to recognize and target tumor cells. Thus, a CAR T-cell therapy combined with a strong activation of DCs could potentially amplify the bystander effects, where the activated CD8 T cells also target neighboring tumor cells that do not express the specific antigen recognized by CAR T-cell. Reader can refer to \cite{lai2020adoptive} for similar experimental study. Similar to previous modelling scenario, DC population, in addition to agents activating DCs, can be linked to bystander cells to create a similar framework. 

So far, our emphasis regarding the bystander dynamics has been based on ODE modelling. However, one can also consider PDE modelling to account for spatial dynamics as well. 
In fact, PDE modelling can be useful for several reasons. Firstly, bystander effect often involves interactions and diffusion of signaling molecules, cytokines, and other mediators within the tumor microenvironment. Thus, PDE models can capture the spatial distribution and diffusion processes more accurately, allowing for a better representation of the complex dynamics occurring in the tumor. Moreover, as outlined in the introduction, solid tumors are characterized by spatial and cellular heterogeneity, with different regions exhibiting distinct characteristics. PDE models can incorporate spatial variations in cell density, antigen expression, and other factors relevant to the bystander effect. For example, in the context of our study, distribution of antigen-positive and negative tumor cells withing the solid tumor bulk would likely have a major impact on the efficacy of the CAR T-cell treatment as well as emergence of bystander effects. However, it is widely recognized that obtaining spatial data is often more challenging than acquiring temporal data in biological and clinical setting. This challenge becomes particularly pronounced when investigating the bystander effect, as outlined in the introduction, due to the complexities involved in setting up and conducting experiments in spatially resolved systems. Consequently, PDE modeling, which incorporates spatial dynamics, may necessitate a more "mechanistic model" approach rather than relying  on data-driven modeling, as exemplified by our ODE model.

Upon the publication of our study, the codes, datasets, and major results of this study will be made publicly available in our GitHub repository.

\newpage
\bibliographystyle{unsrt}
\bibliography{ref-new}

\end{document}